\documentclass[a4paper,11pt]{article}
\pdfoutput=1 

\usepackage{jcappub} 

\usepackage[T1]{fontenc} 

\usepackage{amsmath}
\usepackage{graphicx,epsf,hyperref}

\usepackage{color}
\usepackage{xcolor}

\def\be{\begin{equation}}
\def\ee{\end{equation}}
\def\bea{\begin{eqnarray}}
\def\eea{\end{eqnarray}}
\def\lb{\left(}
\def\rb{\right)}
\def\lbs{\left[}
\def\rbs{\right]}

\def\p{\partial}
\newcommand{\dd}{\mathrm{d}}

\def\Hh{\mathcal{H}}

\def\cs2{c_{\rm{s}}^2}
\def\U0{{\bar U_0}}
\def\12{\frac{1}{2}}

\def\U{{\Upsilon}}

\title{The intrinsic bispectrum of the CMB from isocurvature initial conditions}

 \author{Pedro Carrilho}
 \author{and Karim A. Malik}
 \affiliation{Astronomy Unit, School of Physics and Astronomy, Queen Mary University of London,\\
Mile End Road, London, E1 4NS, UK}

\emailAdd{p.gregoriocarrilho@qmul.ac.uk}
\emailAdd{k.malik@qmul.ac.uk}

\abstract{Non-linear effects in the early Universe generate non-zero bispectra of the cosmic microwave background (CMB) temperature and polarization, even in the absence of primordial non-Gaussianity. In this paper, we compute the contributions from isocurvature modes to the CMB bispectra using a modified version of the second-order Boltzmann solver SONG. We investigate the ability of current and future CMB experiments to constrain these modes with observations of the bispectrum. Our results show that the enhancement due to single isocurvature modes mixed with the adiabatic mode is negligible for the parameter ranges currently allowed by the most recent \emph{Planck} results. However, we find that a large compensated isocurvature mode can produce a detectable bispectrum when its correlation with the adiabatic mode is appreciable. The non-observation of this contribution in searches for the lensing bispectrum from \emph{Planck} allows us to place a new constraint on the relative amplitude of the correlated part of the compensated isocurvature mode of $f_{\rm CIP}=1\pm100$. We compute forecasts for future observations by COrE, SO, CMB-S4 and an ideal experiment and conclude that a dedicated search for the bispectrum from compensated modes could rule out a number of scenarios realised in the curvaton model. In addition, the CMB-S4 experiment could detect the most extreme of those scenarios ($f_{\rm CIP}=16.5$) at 2 to 3-$\sigma$ significance.}

\begin{document}
\maketitle
\flushbottom

\section{Introduction}
\label{sec:intro}

Observations of the cosmic microwave background (CMB) have made it
possible to study the early Universe in detail, as well as to measure
many cosmological parameters with unprecedented precision. The recent
efforts by the \emph{Planck} mission have placed tight constraints on
the nature of the primordial fluctuations by measuring the statistical
properties of the fluctuations in temperature and polarization. In
particular, using the power spectra of the CMB fluctuations, the amplitude of the
power spectrum of adiabatic modes has been measured to be
$\log\left(10^{10} A_{\text{s}}\right)=3.044\pm0.014$ and its spectral
index to be
$n_{\text{s}}-1=0.9649\pm0.0042$~\cite{Akrami:2018odb}. Combining
those temperature and E-mode polarization measurements with those of
B-mode polarization by the BICEP2/Keck collaboration, the
tensor-to-scalar ratio has been constrained to be
$r_{0.05}<0.06$~\cite{Ade:2018gkx}. These contributions have already
ruled out a number of models of inflation and constrained several
others, but have not yet been able to single-out \emph{the}
inflationary model. And, while a possible detection of $r$ by future
experiments such as the Simons Observatory~\cite{Ade:2018sbj},
liteBIRD~\cite{Matsumura:2013aja} or PICO~\cite{Hanany:2019lle} would
severely limit the single-field models of inflation in agreement with
experiment, it would not help in distinguishing those from a
multi-field model of inflation.

One detection that would allow for a distinction between the single and multi-field cases would be that of  isocurvature modes~\cite{Suto:1984aa,Efstathiou:1986pba,Kodama:1986fg,Kodama:1986ud,Mollerach:1989hu,Liddle:1993fq,Peebles:1998nu,Bucher:1999re}. The tightest limits on these non-adiabatic fluctuations come from measurements of the CMB power spectra, which have constrained them to contribute only up to 1\% of the CMB variance~\cite{Akrami:2018odb}. However, large isocurvatures could still exist in the form of a compensated mode~\cite{Gordon:2002gv,Gordon:2009wx} --- a combination between the baryon and cold dark matter isocurvature modes that generates no overall matter isocurvature and leaves the CMB unchanged at the linear level. According to the latest \emph{Planck} results, should this compensated isocurvature mode be uncorrelated with the adiabatic mode, the amplitude of its power spectrum could be as large as $10^{-3}$, i.e. up to 6 orders of magnitude larger that $A_s$. Additionally, since this compensated isocurvature affects the temperature spectrum by broadening its high-$\ell$ peaks in a similar way to lensing, it would help in alleviating the lensing anomaly~\cite{Akrami:2018odb}. Many authors have investigated this mode, by analysing its effects on different probes, in addition to the CMB spectra~\cite{Grin:2011tf,Grin:2011nk,Grin:2013uya,He:2015msa,Munoz:2015fdv,Heinrich:2016gqe,Valiviita:2017fbx,Smith:2017ndr}, such as the 21-cm signal~\cite{Gordon:2009wx}, BBN~\cite{Holder:2009gd}, spectral distortions~\cite{Haga:2018pdl}, magnetogenesis~\cite{Carrilho:2019qlb}, BAO modulation~\cite{Heinrich:2019sxl} and galaxy bias~\cite{Barreira:2019qdl,Hotinli:2019wdp,Barreira:2020lva}. In addition to this compensated mode, even standard isocurvatures with small amplitudes can have large effects on non-linear observables, as shown in Ref.~\cite{Carrilho:2019qlb}, in which the magnetic field produced during pre-recombination times is enhanced on large scales when an isocurvature mode mixes with the adiabatic mode. This is a strong indication that isocurvature modes require further exploration at the non-linear level, which motivates this work on the calculation of the CMB bispectrum generated at second order in the presence of isocurvatures.

Naturally, the main motivation for studying the bispectrum of the CMB is its sensitivity to primordial  non-Gaussianities. By measuring this three-point function, the \emph{Planck} mission has placed constraints on the amplitudes of simple shapes, such as the local shape ($f_{\text{NL}}^{\text{local}}=-0.9 \pm 5.1$), the equilateral shape ($f_{\text{NL}}^{\text{equi}}=-26\pm47$) and the orthogonal shape ($f_{\text{NL}}^{\text{ortho}}=-38\pm 24$) \cite{Akrami:2019izv}. However, given that the primordial signal is small, contributions from secondary effects become important, even substantially biasing the estimation of the primordial signal. This is the case of the lensing contribution from the adiabatic mode to the CMB bispectrum \cite{2009PhRvD..80l3007M,Lewis:2011fk,Lewis:2012tc,2013A&A...555A..82M}, which biases the measurement of $f_{\text{NL}}^{\text{local}}$ by approximately 5 \cite{Akrami:2019izv}. Without knowing about this contribution, a primordial signal would have been wrongly suggested by observations, which highlights the importance of the accurate estimation of its size and shape. Other secondary contributions are also generated, such as the Ricci focusing and redshift modulation effects, which are sourced at recombination. These effects have been estimated analytically~\cite{Creminelli:2004pv,2009JCAP...08..029B,Creminelli:2011sq,Bartolo:2011wb,Lewis:2012tc,Pajer:2013ana} and numerically~\cite{Huang:2012ub,Huang:2013qua,Su:2014tga,pettinari:2013a,Pettinari:2014iha,pettinari:2015a} to have a substantially smaller contribution than the lensing effect, biasing $f_{\text{NL}}^{\text{local}}$ by less than 0.5 with temperature~\cite{pettinari:2013a}, but with an enhanced signal in polarization~\cite{Pettinari:2014iha}. This non-linear signal gives additional information about the early Universe, which could help constraining cosmological parameters or finding modifications to the standard $\Lambda$CDM scenario, such as modified gravity~\cite{Gao:2010um}. Hence, more than a source of error for primordial non-Gaussianity, the bispectrum produced by second-order effects can be used to probe new physics, and this is what we aim to do here with a study of alternative initial conditions.

The plan of this paper is to uncover the effects of isocurvature modes on the intrinsic bispectrum of the CMB. We calculate the bispectrum of the CMB from sources at recombination using the second-order Boltzmann solver SONG~\cite{pettinari:2013a,pettinari:2015a}. While there have been advances in developing the formalism to study more sources and include lensing and time-delay effects numerically~\cite{Fidler:2014zwa,Saito:2014bxa}, we will use the standard version of SONG, modified to include isocurvature initial conditions in addition to adiabatic ones, and will add the effects from lensing, using analytical methods~\cite{Lewis:2012tc}.

The paper is organized as follows. We discuss the standard formalism of second-order perturbation theory and its application to the bispectrum calculation in Section~\ref{sec:bisp}, reviewing the formalism of isocurvature initial conditions. In Section~\ref{sec:res} we present our numerical results for the CMB bispectrum from single isocurvature modes, as well as in mixtures with the adiabatic mode and the interesting case of the compensated isocurvature. Finally, in Section~\ref{sec:conc}, we discuss our results and summarize our conclusions. Throughout the paper, we use units such that $c=\hbar=1$, use Greek letters for spacetime indices and Roman letters for spatial ones and use the mostly plus signature of the metric, $(-,+,+,+)$.

\section{Intrinsic bispectrum of the CMB}
\label{sec:bisp}

The bispectrum of a stochastic field is defined via its three-point function. For this study on CMB fluctuations, the fields of interest are the temperature fluctuations and the E-mode polarization of photons as a function of direction of observation in the sky. These quantities are decomposed into spherical harmonics and the bispectrum we wish to compute is defined by
\be
\left\langle X_{\ell_1m_1}(\vec{x}_0,\tau_0) Y_{\ell_2m_2}(\vec{x}_0,\tau_0) Z_{\ell_3m_3}(\vec{x}_0,\tau_0)\right\rangle=\left(
\begin{array}{ccc} 
 \ell_1 & \ell_2 & \ell_3 \\
 m_1 & m_2 & m_3
\end{array}\right) B_{\ell_1\ell_2\ell_3}^{XYZ}\,,
\ee
where $X$, $Y$ and $Z$ are labelling the fields, which can be $T$, for temperature, or $E$ for E-modes; $\ell$ is the multipole number while $m$ is the azimuthal number. Moreover, we are evaluating the fields at the present position of the earth ($\vec{x}_0$) and at the present time ($\tau_0$). The Wigner-3j symbol was factorized to reveal the angle averaged bispectrum, $B_{\ell_1\ell_2\ell_3}^{XYZ}$, which includes all the information in the bispectrum, given the assumed statistical isotropy.

The bispectrum of a Gaussian field vanishes, thus its existence indicates the distribution of fluctuations is non-Gaussian. Given linear evolution, a bispectrum can only originate via the initial conditions and is thus a probe of primordial non-Gaussianity. However, since the evolution of the Universe is non-linear, a bispectrum will inevitably be generated, even when initial conditions are perfectly Gaussian. It is this contribution that we want to focus on in this paper and for that reason, we now briefly review the concepts of second-order perturbation theory, which will fix our notation and conventions for the remainder of the paper.

We are interested in the evolution of cosmological fluctuations from neutrino decoupling until today, taking special interest in the epoch of recombination. During this time, the species of interest are photons ($\gamma$), electrons ($e$), and protons ($p$), which comprise the tightly coupled baryon-photon plasma, prior to recombination, in addition to cold dark matter (c) and neutrinos ($\nu$). We assume here Einstein's general relativity to describe the geometry through the metric tensor $g_{\alpha\beta}$, obeying
\be 
R_{\alpha\beta}-\frac12 g_{\alpha\beta}R+g_{\alpha\beta}\Lambda=8\pi G T_{\alpha\beta}\,,
\ee
in which $R_{\alpha\beta}$ is the Ricci tensor, $R$ is the Ricci scalar, $G$ is Newton's constant and $T_{\alpha\beta}$ is the stress-energy tensor, which is given by the sum of the stress-energy tensors of all species, labelled by $s$,
\be
T^{\alpha\beta}=\sum_s T_s^{\alpha\beta}\,,
\ee
which are given by
\begin{align}
&T_{\text{c}}^{\alpha\beta}=\rho_{\text{c}} u_{\text{c}}^\alpha u_{\text{c}}^\beta\,,\\
&T_e^{\alpha\beta}=\rho_e u_e^\alpha u_e^\beta\,,\\
&T_p^{\alpha\beta}=\rho_p u_p^\alpha u_p^\beta\,,\\
&T_\gamma^{\alpha\beta}=\frac43\rho_\gamma u_\gamma^\alpha u_\gamma^\beta+\frac13 \rho_\gamma g^{\alpha\beta}+\pi_\gamma^{\alpha\beta}\,,\\
&T_\nu^{\alpha\beta}=\frac43\rho_\nu u_\nu^\alpha u_\nu^\beta+\frac13 \rho_\nu g^{\alpha\beta}+\pi_\nu^{\alpha\beta}\,.
\end{align}
We have here defined the energy densities of each species as $\rho_s$, the 4-velocity vectors as $u_s^\alpha$, and the anisotropic stress tensors for photons and neutrinos as $\pi_\gamma^{\alpha\beta}$ and $\pi_\nu^{\alpha\beta}$, respectively. We are neglecting the pressure and anisotropic stress of electrons, protons and dark matter as those quantities are very small for the times and scales we consider here. Photons and neutrinos are assumed to be relativistic species with an equation of state $P_s=\rho_s/3$. We have written all species in their respective energy frames, as indicated by the lack of an energy flux term, $q^\mu$ in the expressions above.

The particle species evolve according to their Boltzmann equations coupled to the Einstein equations. Particularly relevant is the Boltzmann equation for photons, given by
\begin{equation}
\label{boltzeq1}
\frac{\p f_{\mu\nu}}{\p \tau}+\frac{\p f_{\mu\nu}}{\p x^i}\frac{\dd x^i}{\dd \tau}+\frac{\p f_{\mu\nu}}{\p p}\frac{\dd p}{\dd \tau}+\frac{\p f_{\mu\nu}}{\p n^i}\frac{\dd n^i}{\dd \tau}=\mathcal{C}[f_{\mu\nu}]\,.
\end{equation}
in which $f_{\mu\nu}$ is the Hermitian tensor-valued distribution function of photons, $p$ is the magnitude and $n^i$ is the direction of the 3-momentum of photons in the local inertial frame, and $\mathcal{C}[f_{\mu\nu}]$ is the Thompson collision term, describing the interaction between photons and charged particles. The distribution function can be decomposed in the helicity basis via~\cite{Pitrou:2008hy,Beneke:2010eg,pettinari:2013a,Fidler:2014oda,Pettinari:2014iha,pettinari:2015a}
\be
f^{\mu\nu}=f_{ab}\epsilon_a^{*\,\mu}\epsilon_b^\nu\,,
\ee
with the helicity indices, $a,b$, taking two values, $+,-$ and $\epsilon_a^{\mu}$ representing the polarization state of the photons. Summation over repeated helicity indices is implies here and below. These polarisation vectors are given by
\be
\epsilon_+^i=-\frac{1}{\sqrt{2}}(e^i_\theta+ i e^i_\phi)\,,\ \ \epsilon_-^i=-\frac{1}{\sqrt{2}}(e^i_\theta- i e^i_\phi)\,,
\ee
where $e^i_\theta=\p_\theta n^i$ and $e^i_\phi=\p_\phi n^i/\sin\theta$, in which the direction vector has been expressed in polar coordinates, $\vec n=(\sin\theta\cos\phi,\sin\theta\sin\phi,\cos\theta)$. The Boltzmann equation can also be decomposed into equations for helicity components, $f_{ab}$, which have a similar form to Eq.~\eqref{boltzeq1} when appropriately projected with the polarisation vectors. The components $f_{ab}$ can be decomposed into an intensity part, $f_I$, and E and B-mode linear polarization, $f_E$ and $f_B$, as well as circular polarisation, $f_V$. Here we only consider the intensity and E-mode polarization contributions, since B-modes are expected to have a much smaller amplitude and circular polarisation is not expected to be produced in the standard scenario.

We now perform an expansion around the flat Friedmann--Lema\^{i}tre--Robertson--Walker (FLRW) spacetime, and use Poisson gauge, for which the line element is
\be
\text{d}s^2=a(\eta)^2\lbs-(1+2\phi)\text{d}\eta^2-2S_i\text{d}x^i\text{d}\eta+\lb(1-2\psi)\delta_{ij}+h_{ij}\rb\text{d}x^i\text{d}x^j\rbs\,,
\ee
where $a(\eta)$ is the scale factor, $\phi$ is the perturbation to the lapse, $\psi$ is the curvature fluctuation, $S_i$ is the divergence-free part of the shift and $h_{ij}$ are the transverse and trace-less tensor components. The two scalar potentials as defined in this gauge are equal to the two gauge invariant Bardeen potentials \cite{Bardeen:1980kt}. For simplicity, we will assume that the vector and tensor modes are only non-zero at second order, as we are not considering primordial vector or tensor modes in this study. Our convention for perturbed quantities is $\delta X=\delta X^{(1)}+\delta X^{(2)}/2$. All variables are expanded and decomposed into scalars, vectors and tensors in the standard way, as given in Refs.~\cite{Malik:2008im,Carrilho:2015cma}. 

We define the brightness fluctuations $\Delta_{ab}$ by integrating the perturbation to the photon distribution function with respect to the particle momentum ~\cite{pettinari:2013a,Lewis:2002nc,Beneke:2010eg,Fidler:2014oda},
\begin{equation}
\label{Deltadef}
\delta_{ab}+\Delta_{ab}(\tau,\vec x,\vec n)=\frac{\int{\text{ d}p\, p^3 f_{ab}(\tau,\vec x,p,\vec n)}}{\int{\text{d}p\, p^3 f^{(0)}(\tau,p)}}\,,
\end{equation}
where $f^{(0)}$ is the background distribution function, which is a black body spectrum. 

These fields are typically projected into multipoles by integrating over the direction of the photons with spherical harmonics, resulting in an infinite number of fields, $\Delta_{X\ell m}$ that depend only on spacetime position and not on momentum. Following Refs.~\cite{pettinari:2013a,Fidler:2014oda,Pettinari:2014iha,pettinari:2015a}, we shall denote these projected brightness fluctuations by $\Delta_n$, introducing the composite index $n=X,\ell m$, representing the multipole $\ell m$ of field $X$. We can then write a concise version of the Boltzmann equation after all suitable projections as
\be
\Delta_n'+k\Sigma_{nn'}\Delta_{n'}+\mathcal{M}_n+Q^L_n=\mathcal{C}_n\,,
\ee
in which repeated indices are summed over, we have applied a Fourier transform and we have defined a number of short-hands for different terms in the equation, again following Refs.~\cite{pettinari:2013a,Fidler:2014oda,Pettinari:2014iha,pettinari:2015a}. The matrix $\Sigma_{nn'}$ includes the free-streaming terms, which couple different multipoles; $\mathcal{M}_n$ includes contributions that only depend on the metric tensor; $Q^L_n$ includes quadratic terms involving brightness fluctuations coupled to metric fluctuations, and $\mathcal{C}_n$ is the projection of the collision term, which is given by
\be
\mathcal{C}_n=-|\kappa'|\left(\Delta_n-\Gamma_{nn'}\Delta_{n'}-Q^{\mathcal{C}}_n\right)\,,
\ee
where $\kappa'$ is the Compton scattering rate, the matrix $\Gamma_{nn'}$ encoding the contributions of the gain term and the quadratic terms collected in $Q^{\mathcal{C}}_n$. The detailed form of the terms given here can be found in Ref.~\cite{pettinari:2015a}.

The general Boltzmann equation has thus been recast into an infinite set of coupled differential equations for the different brightness fluctuations, all of which are generated after recombination. Computing their evolution is thus not feasible by solving these differential equations and we are required to use the line-of-sight formula to compute the multipoles at the present time. This is given by 
\be
\Delta_n(\tau_0)=\int^{\tau_0}_0{\text{d}\tau e^{-\kappa(\tau)} j_{nn'}(k(\tau_0-\tau))\mathcal{S}_{n'}(\tau)}\,,
\ee
where the $j_{nn'}$ are combinations of Bessel functions detailed in Ref.~\cite{pettinari:2015a} and $\mathcal{S}_n$ collects the line-of-sight sources, which can be seen from the Boltzmann Equation to be given by
\be
\mathcal{S}_n=-\mathcal{M}_n-Q^L_n+|\kappa'|\left(\Gamma_{nn'}\Delta_{n'}+Q^{\mathcal{C}}_n\right)\,.
\ee
Since the term $\Gamma_{nn'}\Delta_{n'}$ only depends on a few of the multipoles, at first order, this formula allows one to compute only a small number of multipoles required for the accuracy of the sources, which upon integration result in a fast way to calculate an arbitrary number of then at $\tau_0$, as required. At second order, the quadratic terms, $Q^L_n$, which include lensing, time-delay and redshift contributions, include all multipoles once again, making even this approach unsuitable for most purposes. Fortunately, this is not a problem, as alternative methods for calculating those contributions exist, given that Refs.~\cite{Huang:2013qua,Su:2014mga} have introduced clear methods of removing these effects from the line-of-sight sources. In addition, time-delay contributions are found to be negligible~\cite{Huang:2013qua}, lensing contributions are calculable using non-perturbative techniques~\cite{Lewis:2011fk,Lewis:2012tc} and the redshift contribution can be taken into account by making a transformation of variables to~\cite{Huang:2012ub,Fidler:2014oda}
\be
\tilde\Delta_{ab}=\ln(\delta+\Delta)_{ab}\,,
\ee
whose conversion into multipole space can be found in Refs.~\cite{Fidler:2014oda,Pettinari:2014iha}. This resolves all issues and permits a calculation of the second-order evolution of photon perturbations. This is what is done in the Boltzmann solver SONG~\cite{pettinari:2013a,pettinari:2015a}, which makes use of the linear solver CLASS~\cite{Lesgourgues:2011re,Blas:2011rf} to solve for the linear evolution.

In order to compute the bispectrum, it is still required to convert the brightness fluctuations into temperature fluctuations. In this work, we use the same temperature definition used in SONG, the bolometric temperature, whose fluctuations, $\Theta_{ab}$, are related to brightness via~\cite{Fidler:2014oda}
\be
\delta_{ab}+\Delta_{ab}=(\delta+\Theta)_{ab}^4\,,
\ee
which at second order in perturbations is expanded as
\be
\label{DelThe}
\Delta_{ab}=4\Theta_{ab}+6\Theta_{ac}\Theta_{cb}\,.
\ee
Similarly to the distribution function, both the brightness and the temperature fluctuations can be decomposed into intensity and polarisation components. The fields of interest for computing the bispectrum are $T=\Theta_I$ and $E=\Theta_E$. While it is clear in Eq.~\eqref{DelThe}, as well as from the Boltzmann equation, that intensity and polarization components mix together at second order, we reiterate that the contributions from B-modes are assumed to be negligible and those from V-modes are not expected to exist at all in the standard scenario.

SONG has detailed and accurate routines to compute the bispectrum of the CMB, which are detailed in Ref.~\cite{pettinari:2015a}. Beyond the evolution equations and the bispectrum calculation, one is required to specify initial conditions. SONG was built with adiabatic initial conditions. In the next section we review the different non-adiabatic initial conditions which we use to modify SONG.

\subsection{Isocurvature modes}

The solution of the Einstein-Boltzmann system requires the
specification of initial conditions. Different choices of initial
conditions would result in distinct solutions to the evolution
equations, which can be probed with experimental data. An analysis of
the linear Einstein-Boltzmann system~\cite{Bucher:1999re} has led to
the classification of its solution space into regular (growing) and
singular (decaying) modes. While decaying modes have also been studied
with the objective of constraining them via
observations~\cite{Amendola:2004rt,Kodwani:2019ynt,Kodwani:2019qks},
growing modes are certain to exist and we focus only on those. These
growing modes can further be decomposed into an adiabatic and four
isocurvature modes, at first order~\cite{Bucher:1999re}, but only
three isocurvature modes source growing solutions at second
order~\cite{Carrilho:2018mqy}. We define entropy fluctuations of
species $r,s$ as~\cite{Malik:2002jb,Malik:2004tf}
\be
S_{rs}=3{\cal H}\left(\frac{\delta\rho_r}{\rho_r'}-\frac{\delta\rho_s}{\rho_s'}
\right)\,,
\ee
with ${\cal H}$ the Hubble parameter in conformal time, $\delta\rho_s$
the perturbed energy density and $\rho_s$ the energy density in the
background of species $s$. The adiabatic mode is that for which all
entropy fluctuations vanish initially.

The three isocurvature modes are defined instead by the initial
vanishing of the total curvature fluctuation $\zeta$, defined as
\be
\zeta=-\psi-\frac{\delta}{3(1+w)}\,,
\ee
at first order, where $\delta=\delta\rho/\rho$ is the density contrast and $w=P/\rho$ the equation of state parameter. Each isocurvature mode is defined by having one non-zero entropy fluctuation relative to photons, $S_{s\gamma}$, and are named the baryon (BI), cold dark matter (CDI) and neutrino density isocurvature (NI) modes, depending on which entropy fluctuation is non-zero. A more detailed definition of isocurvature modes is included in Ref.~\cite{Carrilho:2018mqy}, which is also consistently extended to second order. 

A particularly simple isocurvature mode is the compensated
isocurvature mode, in which the total matter entropy vanishes due to
the cancellation between the non-vanishing entropy perturbations of
baryons and dark matter. This mode therefore obeys
\be
\rho_c S_{c\gamma}+\rho_b S_{b\gamma}=0\,.
\ee
This cancellation means that this mode does not evolve and all other fluctuations vanish, at the linear level, in the approximation that baryons are pressureless. When this mode is present along with the adiabatic mode it can generate evolution at second order, due to the mixing of both modes. This can be easily understood by the fact that this compensated mode is effectively a space-dependent variation of the baryon-to-dark matter ratio. When present on large scales, its effect on the CMB is to modulate the high-$\ell$ peaks in the temperature power spectrum in a similar way to lensing. This happens due to the effective averaging of multiple regions of the sky with differing baryon-to-dark matter ratios, which slightly broadens and dampens the small-scale peaks of the spectrum. This a clear sign of a coupling between large and small scales and therefore, it is expected that this compensated mode generates substantial non-Gaussianity, especially of the local type, as we will see below.

The most general solution of the Einstein-Boltzmann system at first order is a linear combination of all modes with different amplitudes. At second order, this results in quadratic combinations of different modes, thus implying that they mix and create novel effects that do not exist at the linear level. This general solution for any variable $X(\mathbf{k},\tau)$ can be written as
\be
X(\mathbf{k},\tau)=\mathcal{T}^a_X(\mathbf{k},\tau) I_a(\mathbf{k})+\int \frac{\text{d}^3k_1\text{d}^3k_2}{(2\pi)^3}\delta^{(3)}(\mathbf{k}-\mathbf{k_1}-\mathbf{k_2})\mathcal{T}^{ab}_X(\mathbf{k},\mathbf{k_1},\mathbf{k_2},\tau) I_a(\mathbf{k_1})I_b(\mathbf{k_2})\,,
\ee
where $I_a$ denotes the variable defining the mode, taking values from the set $\{\zeta,S_{b\gamma},S_{c\gamma},S_{\nu\gamma}\}$, $\mathcal{T}^{a}$ and $\mathcal{T}^{ab}$ are transfer functions at first and second order, respectively, and repeated indices are summed over. 

The initialization of the Boltzmann solver SONG is made via approximate initial solutions to the transfer functions above, obtained via an expansion in powers of $\tau$. This is detailed in Ref.~\cite{Carrilho:2018mqy}, which includes all the approximate initial solutions in the tight coupling limit, which we use here.

SONG includes also the tight coupling approximation (TCA) going up to first order in interaction time, $1/\kappa'$, for the adiabatic case. We have modified this part of SONG to account for non-adiabatic contributions to the velocity slip, $\Delta v_{b\gamma}=v_b-v_\gamma$, which is now given by
\begin{align}
\Delta v_{b\gamma\, m}=\frac{1}{\kappa'}\frac{R}{1+R}&\left(\Hh v_{b\gamma\,m}-S_m\delta_{|m|1}-\frac{\delta_\gamma}{4}\delta_{m0}+2\Hh v_{b\gamma\,m}\delta_b-2 \left[v_{b\gamma\, m}\left(\frac{\delta_\gamma}{4}-S_{b\gamma}\right)\right]'\right.\nonumber\\
&\left.+\frac{Q^L_{b\, 1m}}{3 k}-\frac{Q^L_{\gamma I\,1m}}{4 k}\right)+\frac{1}{4k} Q^C_{I\,1m}+v_{b\gamma\,m}\left(\frac{\delta_\gamma}{4}-S_{b\gamma}\right)\,,
\end{align}
where the dependence on the azimuthal number $m$ has been made explicit, $R=\rho_b/\rho_\gamma$, $S_m$ is the multipole expansion of the vector potential $S^i$, $Q^L_{b\, 1m}$ is the quadratic part of the dipole of the Liouville term for baryons, which can be obtained from the conservation of stress-energy for the baryon fluid, and $v_{b\gamma}$ is the average velocity of the baryon-photon plasma, which is related to the initial intensity dipole via $\Delta_{I\,1m}=-4 k v_{b\gamma\,m}(1+\delta_\gamma)$~\cite{pettinari:2015a}. All quadratic expressions are convolutions in Fourier space, which have been simplified here for brevity. The contributions from the baryon entropy, $S_{b\gamma}$, are made explicit above and had not all been included in previous versions of SONG.

For all modes, we assume that their initial fluctuations are Gaussian and are thus solely determined by their power spectrum
\be
\langle I_a(\mathbf{k}) I_b(\mathbf{k}')\rangle=(2\pi)^3\delta(\mathbf{k}+\mathbf{k}')P_{ab}(k)\,,
\ee
which we assume has power-law form given by
\be
P_{ab}(k)=A_{ab} \frac{2\pi^2}{k^3} \left(\frac{k}{k_*}\right)^{n_{ab}}\,,
\ee
where $A_{ab}$ is the amplitude, $n_{ab}$ is the spectral index and $k_*=0.05$ Mpc$^{-1}$ is the CMB pivot scale. For the adiabatic mode, $A_{\zeta\zeta}=A_s$ and $n_{\zeta\zeta}=n_s-1$, as typically defined in the literature. For isocurvatures, we quote directly $n_I\equiv n_{II}$, not including the factor of $-1$ that is conventional for the adiabatic mode. For $a\neq b$, it is also useful to define the correlation angle,
\be
\cos\theta_{ab}=\frac{P_{ab}}{\sqrt{P_{aa}P_{bb}}}\,,
\ee
which takes values between -1 and 1, with the extremes describing full anti-correlation and full correlation, respectively. 

Given the definitions above, the bispectrum generated at second order is given, schematically, by the following expression
\begin{align}
\label{Bisp_transfers}
B_{\ell_1\ell_2\ell_3}^{XYZ} = \sum_{m,L_1,L_3}\int_{k_1,k_2,k_3,r} K_{L_i,\ell_j,m}(k_i,r)\mathcal{T}_{X_{\ell_1 m}}^{ab}(k_1,k_2,k_3) \mathcal{T}_{Y_{\ell_2 0}}^{c}(k_2) &\mathcal{T}_{Z_{\ell_3 0}}^{d}(k_3) P_{ac}(k_2) P_{bd}(k_3)\nonumber\\
  &+\ \text{perms.}
\end{align}
We have collected a number of factors in the coefficient $K_{L_i,\ell_j,m}(k_i,r)$, which depends on all wave-numbers, $k_i$, and multipole numbers, $\ell_i$, as well as on auxiliary multipole numbers $L_1$, $L_3$. Its expression can be easily determined from Eq.~(6.36) in Ref.~\cite{pettinari:2015a}, which is the final formula for the bispectrum used by SONG. Note additionally that only the second-order transfer function depends on the azimuthal index, $m$, since non-scalar contributions are only generated at second order. While this number should be summed between $-\infty$ and $+\infty$, in practice, one accounts for tensor modes already at $|m|=2$ and all higher $|m|$ are difficult to generate. In addition, at least in the adiabatic case, it has been shown that including up to $|m|=3$ only modifies the bispectrum by up to 3\% relative to including only $m=0$~\cite{Pettinari:2014iha,pettinari:2015a}. We also test this for the isocurvature case, motivated by the fact there exist additional ways to produce non-scalar modes when isocurvatures mix with adiabatic modes, as exemplified in the enhanced production of magnetic fields~\cite{Carrilho:2019qlb}.

\subsection{Fisher Forecasts}

To establish the observability of the effects of isocurvatures, we use a Fisher matrix approach. We use the amplitude of the possible bispectra as the observables of interest and thus, the elements of the Fisher matrix are simply given by
\be
F^{(a),(b)}=f_{\rm{sky}}\sum_{X_i,Y_j}{\sum_{2\leq \ell_1\leq \ell_2\leq \ell_3}^{\ell_{\rm{max}}}{\Delta_{\ell_1\ell_2\ell_3}^{-1}B^{(a),X_1X_2X_3}_{\ell_1\ell_2\ell_3}(C_{\rm{tot}}^{-1})_{\ell_1}^{X_1Y_1}(C_{\rm{tot}}^{-1})_{\ell_2}^{X_2Y_2}(C_{\rm{tot}}^{-1})_{\ell_3}^{X_3Y_3}B^{(b),Y_1Y_2Y_3}_{\ell_1\ell_2\ell_3}}}\,,
\ee
where $X_i$ and $Y_i$ represent the fields $T$ or $E$, $\Delta_{\ell_1\ell_2\ell_3}$ is a combinatorial factor, being 6 for the equilateral case, 2 for the isosceles case and 1 otherwise. $C_{\rm{tot}}^{XY}$ is the total spectrum including a noise contribution, and the product of the three inverse spectra constitute the covariance of the bispectra in the Gaussian approximation. In addition to the intrinsic bispectrum, we will also consider the lensing bispectrum, as well as the bispectrum generated by primordial local non-Gaussianity, with the aim of estimating how those measurements are biased by ignorance of the intrinsic signal. The signal-to-noise ratio for the intrinsic component is given by $S/N=\sqrt{F^{\rm{int,int}}}$, while the bias in the measurement of bispectrum $(a)$ by the intrinsic bispectrum is $\Delta f_{\rm{NL}}^{(a)}=F^{(a) \rm{int}}/F^{(a)(a)}$.

It should be noted that using a Gaussian approximation for the covariance is not always accurate. In particular it is well known that the non-Gaussianity generated by lensing is large enough to alter the covariance of the bispectrum, given its large connected trispectrum~\cite{Lewis:2011fk,Coulton:2019odk}. It is therefore conceivable that additional contributions to the covariance exist in the case of non-Gaussianity generated by isocurvatures. However, these require the calculation of the 6 and 4-point functions, which is not currently possible in general. Ref.~\cite{Pettinari:2014iha} estimates the effects from lensing to degrade the signal-to-noise by approximately $\sqrt{2}$. For isocurvature modes that generate non-Gaussianity at similar levels to the lensing effects, we estimate the degradation of $S/N$ to be similar. Therefore, in several places below, we refer to the total degradation due to all these effects as being a factor of 2 decrease of the signal to noise. It should be noted, however, that de-lensing techniques such as those described by Ref.~\cite{Coulton:2019odk}, could reduce the signal covariance due to lensing, which would make the error in our forecasts smaller. In general, therefore, our forecasts could be interpreted in the context of the detection of the effects of isocurvatures away from zero, should de-lensing be very effective. 

\section{Results}
\label{sec:res}

We compute the bispectrum for different initial conditions, starting with a review of the adiabatic case. As described above, the bispectrum includes contributions that are not computed by SONG. The time delay contribution is negligible and the lensing bispectrum is given by \cite{Lewis:2011fk,Lewis:2012tc}
\be
B_{\ell_1\ell_2\ell_3}^{\text{lens},\,X_1X_2X_3}=f^{X_3}_{\ell_3\ell_1\ell_2}C^{X_1\psi}_{\ell_1}\tilde C^{X_2X_3}_{\ell_2}+\text{perms}\,,
\ee 
where $C_\ell^{XY}$ is the angular power spectrum for fields $X,Y=T,E$, $C^{X\psi}$ is the cross-power spectrum between $X=T,E$ and the lensing potential $\psi$, and $f^{X_i}_{\ell_1\ell_2\ell_3}$ is given by
\begin{align}
f^{T}_{\ell_1\ell_2\ell_3}&=\frac12[\ell_2(\ell_2+1)+\ell_3(\ell_3+1)-\ell_1(\ell_1+1)]\\
f^{E}_{\ell_1\ell_2\ell_3}&=\frac12[\ell_2(\ell_2+1)+\ell_3(\ell_3+1)-\ell_1(\ell_1+1)]\left(
\begin{array}{ccc} 
 \ell_1 & \ell_2 & \ell_3 \\
 2 & 0 & -2
\end{array}\right) \left(
\begin{array}{ccc} 
 \ell_1 & \ell_2 & \ell_3 \\
 0 & 0 & 0
\end{array}\right)^{-1}\,,
\end{align}
where the brackets represent Wigner-3j symbols, as above. For the case of temperature, the biggest contribution to $C^{T\psi}$ comes from the integrated Sachs-Wolfe effect, whereas for E-mode polarization it is the effects of reionization that affect $C^{E\psi}$ the most. The bispectra generated by these lensing effects peaks in the squeezed configuration and is the largest contribution to the full bispectrum generated at second order, having been detected by the Planck collaboration~\cite{Akrami:2019izv}, ruling out its non-existence at approximately $3.5\sigma$ confidence.\footnote{Note that this $3.5\sigma$ significance is calculated under the hypothesis of this signal not existing and for that reason assumes a Gaussian covariance. As mentioned at the end of the previous section, a more realistic covariance would increase the errors and therefore reduce the constraining power on the amplitude of this lensing contribution to below $3\sigma$.} This is clearly an important effect to account for in searches for primordial non-Gaussianity, as it can cause a bias on local non-Gaussianity of $\Delta f_{\text{NL}}^ {\text{loc}}\approx 5$ when ignored. 

\begin{figure}[h]
    \centering
    \includegraphics[scale=0.49]{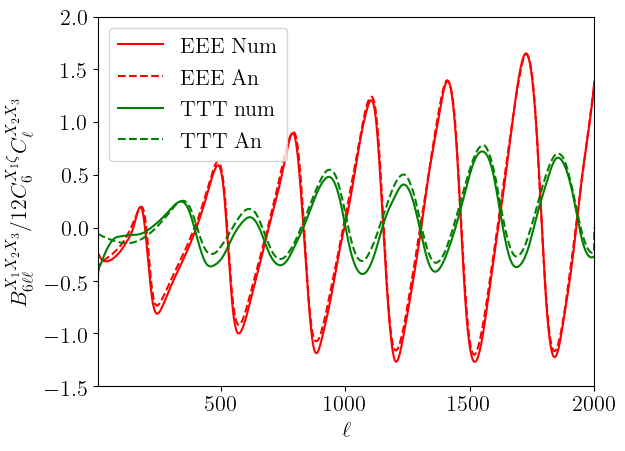}
		\includegraphics[scale=0.49]{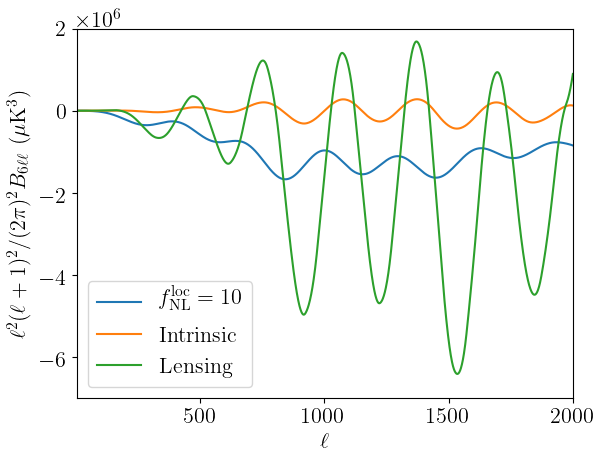}
    \caption{Intrinsic bispectrum of the CMB for adiabatic initial conditions in the squeezed limit. Left --- Comparison between the numerical result and the analytical expression in Eq.~\eqref{RicciFocus} for $\ell_1=6$, $\ell_2=\ell_3=\ell$. Right --- Comparison between the lensing and intrinsic contributions to the TTT bispectrum as well as to a example local shape for $f_{\rm{NL}}^{\rm{loc}}=10$, for $\ell_1=6$, $\ell_2=\ell$, $\ell_3=\ell-6$ (for which lensing is maximal).}
    \label{AdAn_comp}
\end{figure}

The remaining contribution to the bispectrum has been calculated numerically before, in the adiabatic case, both using SONG~\cite{pettinari:2013a,Pettinari:2014iha,pettinari:2015a} and other Boltzmann solvers~\cite{Huang:2012ub,Huang:2013qua,Su:2014tga}. This contribution is substantially smaller than that from lensing, in the adiabatic case. It also peaks in the squeezed configuration, but has a different shape. In the squeezed limit, a simple analytical formula can be found for the bispectrum in the adiabatic case, by considering the contribution of Ricci focusing. Including also the contribution from redshift modulation, the analytical expression is given by~\cite{Creminelli:2004pv,Creminelli:2011sq,Bartolo:2011wb,Lewis:2012tc,Pettinari:2014iha}
\be
\label{RicciFocus}
B_{\ell_1\ell_2\ell_3}^{\text{sq},\,X_1X_2X_3}=C^{X_1T}_{\ell_1}\left(\delta_{X_2T} C^{X_2 T}_{\ell_2}+\delta_{X_3T} C^{X_3 T}_{\ell_3}\right)-\frac12 C^{X_1\zeta}_{\ell_1}\left(\frac{1}{\ell_2}\frac{d (\ell_2^2 C_{\ell_2})}{d \ell_2} +\frac{1}{\ell_3}\frac{d (\ell_3^2 C_{\ell_3})}{d \ell_3}\right)\,,
\ee
where $\ell_1\ll\ell_2,\ell_3$ and $C^{X\zeta}_\ell$ is the angular spectrum between $X=T,E$ and the curvature perturbation $\zeta$ at recombination. In Fig.~\ref{AdAn_comp}, we show the agreement between this formula and the numerical result from SONG, as well as the comparison between the lensing and the intrinsic contributions, for temperature, also in the squeezed limit. In the right-hand plot in this figure, we show also the bispectrum for the local shape in the squeezed limit, which, for a general value of $f_{\rm NL}^{\rm loc}$ is given by
\be
B_{\ell_1\ell_2\ell_3}^{\text{loc},\,X_1X_2X_3}\approx \frac65 f_{\rm NL}^{\rm loc}C^{X_1\zeta}_{\ell_1}\left(C^{X_2 X_3}_{\ell_2}+C^{X_2 X_3}_{\ell_3}\right)\,,
\ee
where $\ell_1\ll\ell_2,\ell_3$. In the left-hand plot of Fig.~\ref{AdAn_comp}, we divide the bispectrum by this local shape to get an $O(1)$ result. However, this modifies the oscillations somewhat, so in the right-hand plot and in all other figures below, we choose to multiply instead by $(\ell(\ell+1)/2\pi)^2$, which is the square of the overall scaling of the angular power spectrum, in analogy to what is typically done for plotting the power spectrum.

This intrinsic signal is difficult to observe, given its small amplitude. For \emph{Planck}, the forecasted signal-to-noise was only $S/N\approx0.5$, but future experiments with increased precision in polarization may detect this signal, as an ideal experiment is forecasted to get $S/N\approx 4$ for $\ell_{\text{max}}=4000$~\cite{Pettinari:2014iha}.

\subsection{Single isocurvature}

We now show our results for single isocurvature modes, studying all three kinds of density isocurvatures. We begin with the case in which only the isocurvature mode is present, without the influence of the adiabatic mode. This case is somewhat ``academic'', since it is known that the adiabatic mode is present from its detection in the power spectrum. Still, some information can be obtained from this analysis. For concreteness, we assume the amplitude of the isocurvature mode in question to be the same as that measured for the adiabatic mode, i.e. $A_{I}=A_s$ and that its spectral index is $n_{I}=0$.

\begin{figure}[h]
    \centering
    \includegraphics[scale=0.49]{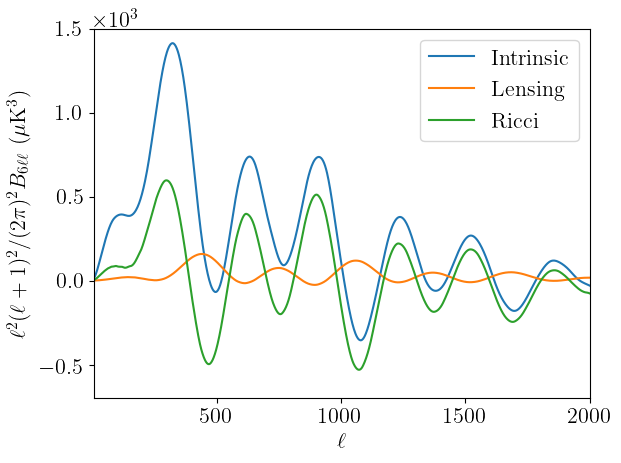}
		\includegraphics[scale=0.49]{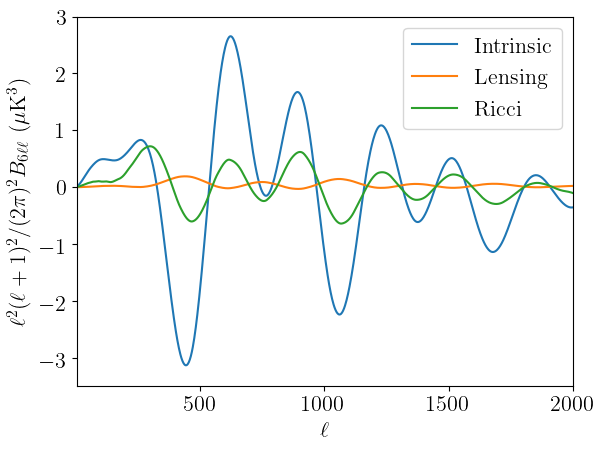}
		\includegraphics[scale=0.49]{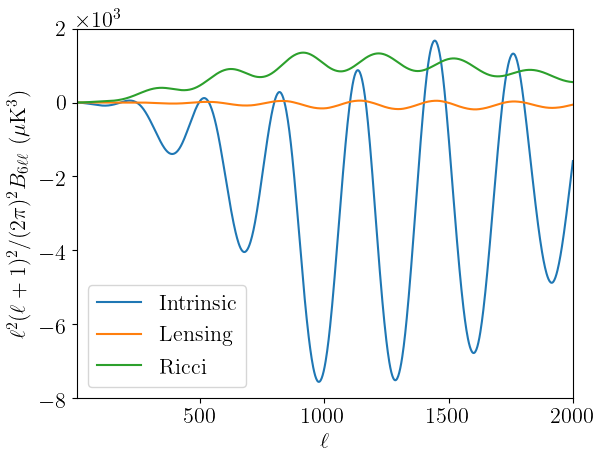}
    \caption{Intrinsic TTT bispectrum of the CMB for single isocurvature initial conditions in the squeezed limit, compared to the bispectrum due to lensing as well as the expectation from the Ricci focusing formula, Eq.~\eqref{RicciFocus}. Top left --- Dark matter isocurvature. Top right --- Baryon isocurvature. Bottom --- Neutrino density isocurvature. The lensing contributions are show in the configuration $\ell_1=6$, $\ell_2=\ell$, $\ell_3=\ell-6$.}
    \label{Lonely_Iso}
\end{figure}

Fig.~\ref{Lonely_Iso} shows the bispectrum in the squeezed limit for all three cases. While very different in all cases, it is clear that the peaks of the bispectrum are far smaller for isocurvatures than for the adiabatic case. For these fixed amplitudes of the modes, the largest bispectrum is for the neutrino isocurvature, which is approximately two orders of magnitude smaller than the adiabatic case. It is therefore infeasible to ever detect these intrinsic contributions to the bispectrum, even if isocurvatures were present with the same amplitude of the adiabatic mode. 

Additionally, the lensing contributions are sub-dominant in all three cases, contrary to what occurred for the adiabatic case. This is explained by the fact that isocurvature initial conditions have negligible gravitational potentials initially, which grow during radiation domination, but always to a smaller level than in the adiabatic case. Both lensing and ISW, being sourced by the sum of the potentials $\phi$ and $\psi$, are thus smaller in the isocurvature case, with the consequence of a smaller lensing bispectrum. Therefore, this is also not a useful signal for detecting isocurvature modes.

The result from Eq.~\eqref{RicciFocus}, the so-called Ricci focusing is also shown in Fig.~\ref{Lonely_Iso}, demonstrating that this formula no longer explains the intrinsic bispectrum accurately. This is expected given that the Ricci focusing formula was found for the adiabatic case and should not be applied to this situation in which the curvature perturbation is time-dependent on large scales due to the presence of entropy fluctuations.

We can also see from Fig.~\ref{Lonely_Iso} that the results for the two
matter isocurvatures are not proportional to each other, contrary to what
happens with the power spectrum. The reason for this is specially
clear in the squeezed limit, in which small scales are modulated by a
large scale field. In the case of matter isocurvatures, that large
scale field is effectively the local matter density corresponding to
the mode in question. The modulation from the density of baryons is
very different from that from the dark matter density, which generates
these very different bispectra.

We have limited our analysis to the temperature case, in the squeezed limit, as that is where these bispectra peak. The situation with polarization is similar, as well as with other configurations. In all cases, these pure isocurvature contributions are too small to ever be detected, given current constraints on their amplitude.
\\

We now move to the case in which one isocurvature mode mixes with the adiabatic mode. Again, we make the same assumptions for the amplitude and spectral index of isocurvatures as above, i.e. $A_{I}=A_s$ and $n_{I}=0$. We also begin by considering uncorrelated modes, i.e. with $\cos\theta_{\zeta I}=0$.

\begin{figure}[h]
    \centering
    \includegraphics[width=0.49\textwidth]{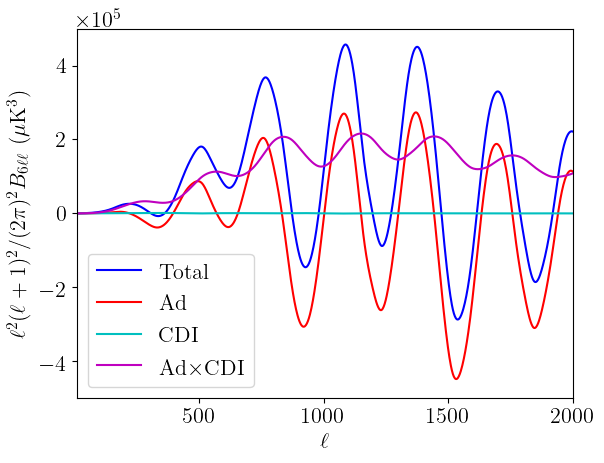}
		\includegraphics[width=0.49\textwidth]{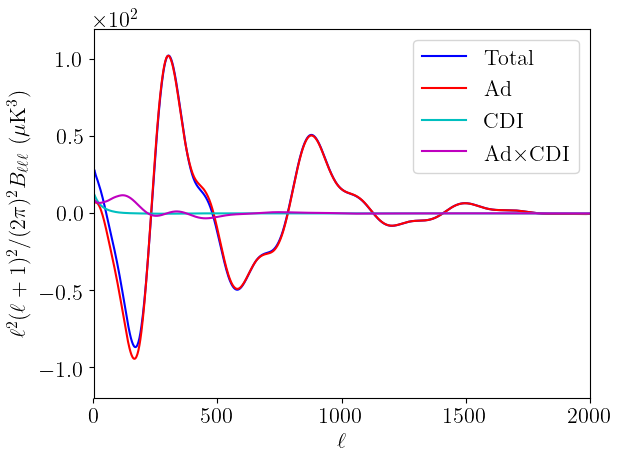}
		\includegraphics[width=0.49\textwidth]{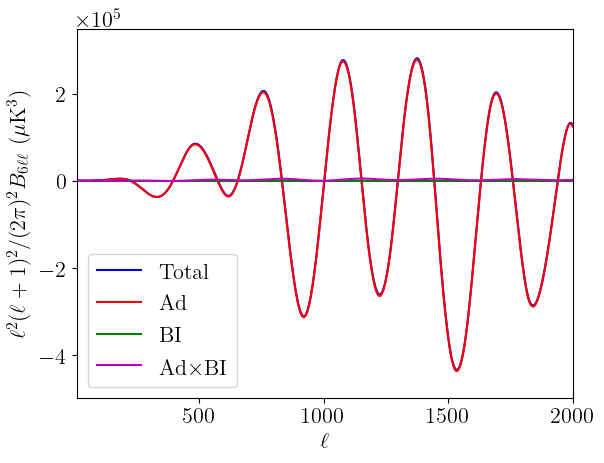}
		\includegraphics[width=0.49\textwidth]{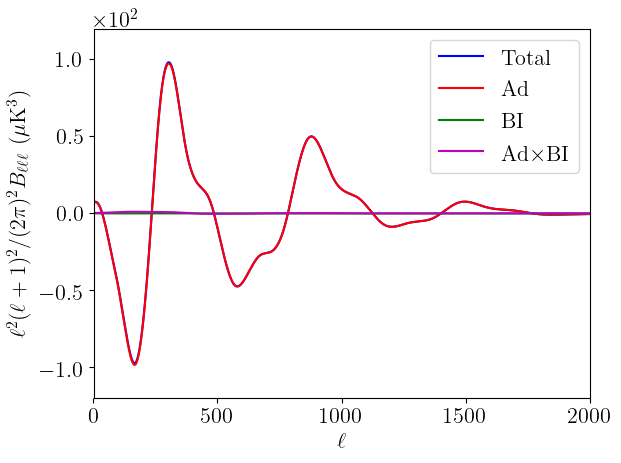}		
		\includegraphics[width=0.49\textwidth]{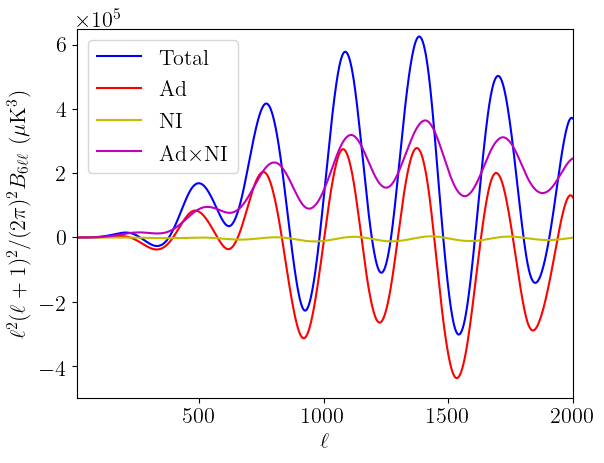}
		\includegraphics[width=0.49\textwidth]{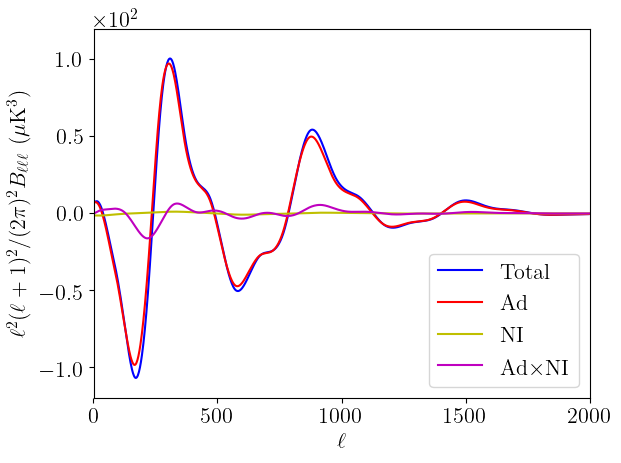}
    \caption{The intrinsic TTT bispectrum of the CMB for single isocurvature modes mixed with adiabatic mode, in the squeezed (left) and equilateral (right) configurations, including both pure contributions, the contributions from mode mixing and the total result. Top --- Dark matter isocurvature (CDI). Middle --- Baryon isocurvature (BI). Bottom --- Neutrino density isocurvature (NI).}
    \label{Mixed_Iso_Sq}
\end{figure}

We show our results for the TTT bispectra in Fig.~\ref{Mixed_Iso_Sq}. We confirm once more that the pure contributions (BI, CDI or NI) are sub-dominant relative to the adiabatic contribution. However, the mixed contributions are substantial and do increase the signal in a distinctive way in the squeezed configuration. This contribution is very similar to that generated by local primordial non-Gaussianity and it is clear that ignoring it biases the primordial measurement. The effect is similar in all cases, in spite of the low amplitude in the baryon isocurvature case, and is a clear probe of the amplitude of the isocurvature mode.

\begin{figure}[h]
    \centering
    \includegraphics[width=0.5\textwidth]{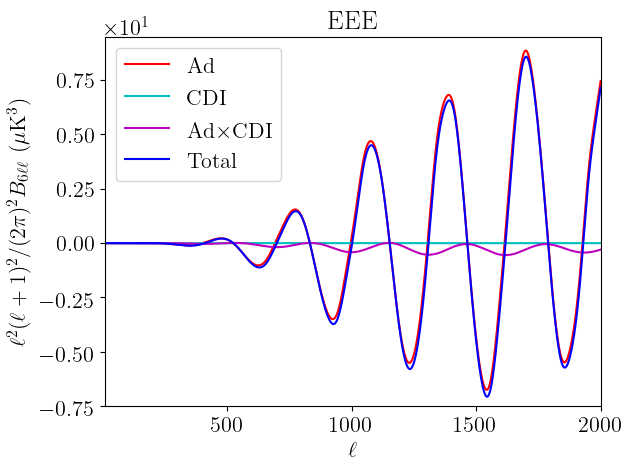}
		\includegraphics[width=0.48\textwidth]{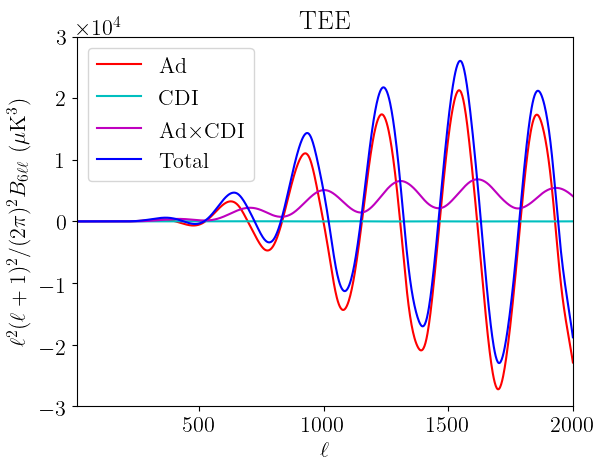}
		\caption{Intrinsic bispectrum of the CMB for dark matter isocurvature mode mixed with adiabatic mode, in the squeezed configuration compared to the contribution from the adiabatic mode for different combinations of $E$ polarization and temperature $T$ given by $X_1X_2X_3$ above each plot, with $X_1$ corresponding to the squeezed field.}
    \label{CDI_pol_Sq}
\end{figure}

The plots for the equilateral configuration show only very small contributions from the isocurvature modes and only on large scales, with the mixed contributions being the largest, but still rather small compared to the adiabatic case. For polarization, the contributions from isocurvatures are smaller, including in the squeezed configuration, as shown in Fig.~\ref{CDI_pol_Sq} for the dark matter isocurvature. Therefore, for bispectra including a mixture of temperature and polarization, some effects of isocurvature modes are present, but for the $EEE$ bispectrum, those effects are negligible.

\begin{figure}[h]
    \centering
    \includegraphics[width=0.485\textwidth]{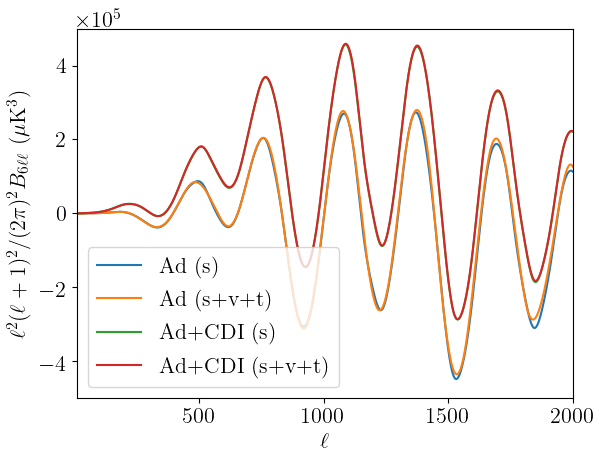}
		\includegraphics[width=0.505\textwidth]{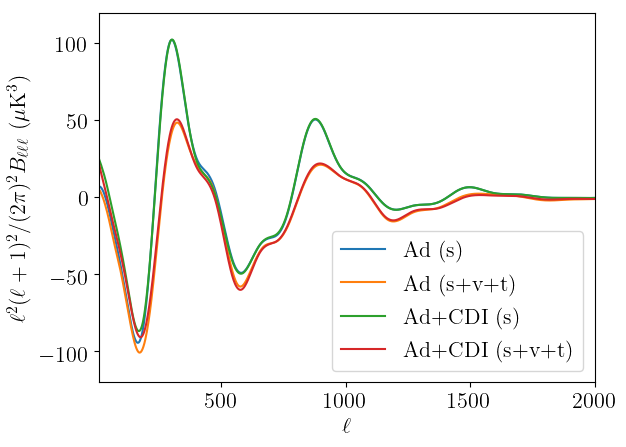}
		\caption{Intrinsic TTT bispectrum of the CMB for dark matter isocurvature mode mixed with adiabatic mode demonstrating the effect of adding vector and tensor modes in the squeezed (left) and equilateral (right) configurations.}
    \label{CDI_vec_ten}
\end{figure}

Next, we explore the effects of vector and tensor modes, which are generated at second-order and can be enhanced by isocurvature modes~\cite{Christopherson:2009bt,Christopherson:2010ek,Carrilho:2019qlb}. We find that adding these modes influences the bispectrum in non-squeezed configurations, as seen in Fig.~\ref{CDI_vec_ten} and is largest for the adiabatic contribution, with the isocurvature contribution changing only negligibly when vectors and tensors are added. Our results for polarization lead to similar conclusions, with even smaller effects. Given that the squeezed configuration is dominant and the effects there are negligible, the addition of non-scalar modes only increases the signal-to-noise of the bispectrum by up to 3\%, in line with what was found for the adiabatic case~\cite{Pettinari:2014iha}.

\begin{figure}[h]
    \centering
    \includegraphics[width=0.50\textwidth]{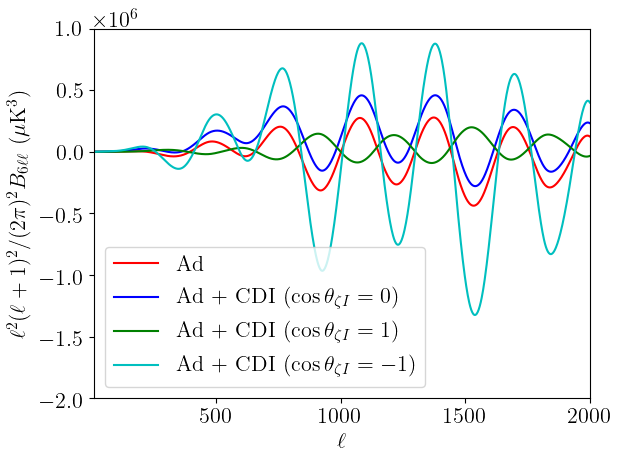}
		\includegraphics[width=0.49\textwidth]{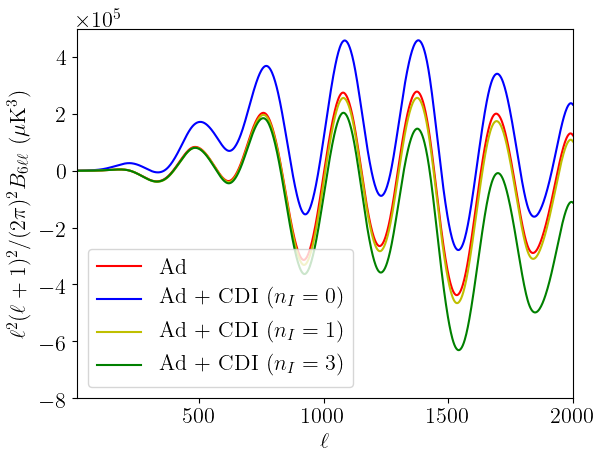}
		\caption{Intrinsic TTT bispectrum of the CMB for dark matter isocurvature mode mixed with adiabatic mode, in the squeezed configuration, varying correlation angle, $\cos\theta_{\zeta I}$, (left) and spectral index, $n_I$ (right).}
    \label{CDI_var_n_corr}
\end{figure}

Moving away from the simplest scenario of a scale-invariant, uncorrelated isocurvature mode can give rise to further modifications in the bispectrum. As seen in the left plot of Fig.~\ref{CDI_var_n_corr}, a positive correlation angle reduces the bispectrum, while a negative correlation greatly enhances it approximately by a factor of 3. In the case of $\cos\theta_{\zeta I}=1$ this reverts the phase of the bispectrum in this configuration and lowers its amplitude below that of the adiabatic mode alone. In other configurations, it is only the large scales (low $\ell$) which are affected by the correlation, changing in the same direction as the squeezed case. Similar effects are seen already in the power spectrum, for which varying the correlation angle greatly modifies the spectrum on large scales, reducing it for positive correlation and increasing it for negative, in parallel with what occurs in the bispectrum.

Regarding the variation of the spectral index, $n_I$, the effects are smaller, but also parallel to those occurring in the power spectrum in similar conditions. We see on the right plot of Fig.~\ref{CDI_var_n_corr}, that the case $n_I=1$ nearly removes the effect of the isocurvature mode, while increasing the spectral index further increases the effect in the negative direction. For the equilateral case, the effects of varying $n_I$ are smaller, but also move the bispectrum closer to the adiabatic case. Note that $n_I=1$ is currently the least constrained scenario by the power spectrum analysis of \emph{Planck}. It is clear from here that the bispectrum also would have trouble distinguishing that case from the adiabatic case, should it be included in future analysis.

Many of the results shown above show some promise as probes of the amplitude of isocurvatures. However, two points need to be made regarding the observability of those bispectra. First, the effects that increase the signal shown above also increase its covariance by enhancing the power spectrum in similar ways. In most cases, therefore, this results in a lower $S/N$ for the total intrinsic bispectrum than expected, often being even lower than for the adiabatic case, which would compromise the potential of these effects to test isocurvatures. The other issue is the fact that the amplitudes for the isocurvature modes used as examples above are substantially larger than the constraints from the latest \emph{Planck} analysis of the power spectrum~\cite{Akrami:2018odb}, except for the baryon isocurvature. With realistic amplitudes, these effects are much smaller, indeed being similar to those shown for the baryon isocurvature case. This severely compromises the ability of any future survey to use the intrinsic bispectrum to improve constraints on single isocurvatures. In spite of this, a far more promising avenue is the exploration of compensated isocurvatures, which avoid both issues and are discussed in the next section.

\subsection{Compensated isocurvature}

We now move to the case of the compensated isocurvature, often labelled CIP, for Compensated Isocurvature Perturbation~\cite{Gordon:2002gv,Gordon:2009wx,Grin:2011tf,Grin:2011nk,Grin:2013uya,He:2015msa,Munoz:2015fdv,Heinrich:2016gqe,Valiviita:2017fbx,Smith:2017ndr}. Given the set of initial conditions defined above, this mode can be seen as a combination between the two matter isocurvature modes, which compensate each other in the way described above. The motivation for specifying such a combination is that, contrary to other isocurvatures, this mode is fairly unconstrained and the bispectrum can be used to probe it further, given its sensitivity to second-order effects.

\begin{figure}[h]
    \centering
		\includegraphics[width=0.49\textwidth]{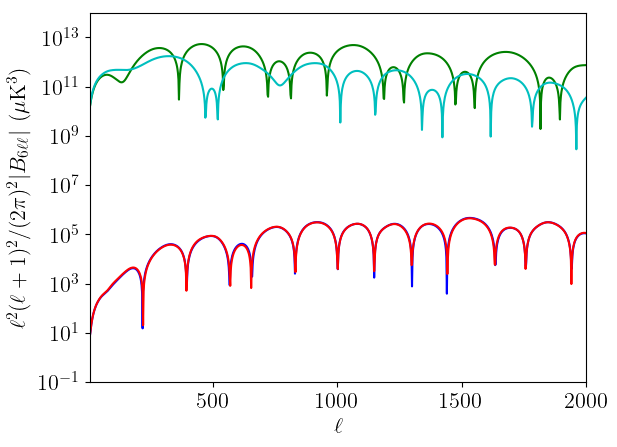}
    		\includegraphics[width=0.49\textwidth]{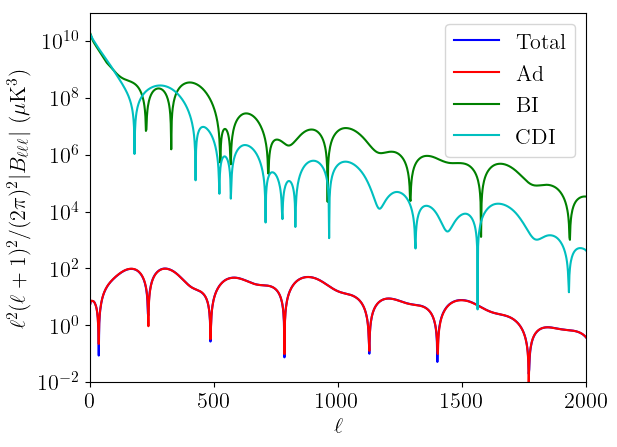}
    \caption{Absolute value of the TTT bispectrum for a compensated isocurvature with $A_{\rm{CIP}}=10^6 A_s$ uncorrelated with the adiabatic mode, compared to the result for single baryon and cold dark matter isocurvatures contributions, in the squeezed (left) and equilateral configurations (right).}
    \label{CIP_uncorr}
\end{figure}

The correlation between the compensated mode and the adiabatic mode is crucial for its effect on the bispectrum, with an uncorrelated compensated isocurvature giving no contribution at all. This can be seen most clearly from Eq.~\eqref{Bisp_transfers}. It is known that these modes have no effect on the CMB at the linear level, implying that if either $c,d=\text{CIP}$, the transfer functions are zero and that contribution to the bispectrum vanishes. One must therefore have that both linear transfer functions are for the adiabatic mode (i.e.~$c,d=\zeta$), which implies the only way for compensated modes to contribute is if one of the spectra in Eq.~\eqref{Bisp_transfers} is a non-zero correlation between the adiabatic and CIP modes (i.e.~$P_{\zeta\,\rm{CIP}}\neq0$). While the existence of this correlation is well motivated, since it appears in curvaton scenarios, it also implies that the bispectrum is insensitive to the uncorrelated CIP. This fact can actually be used to test SONG, to confirm that all cancellations between baryon and dark matter modes are correctly taken into account. This is shown in Fig.~\ref{CIP_uncorr}, in which we plot the intrinsic bispectrum in both the squeezed and equilateral configurations for that case, where we chose $A_{\rm{CIP}}=10^6 A_s$ for illustrative purposes. This shows also the individual contributions from pure matter isocurvatures (BI and CDI) and then demonstrates that all of them cancel out and the total result is indistinguishable from the adiabatic one.

\begin{figure}[h]
    \centering
    		\includegraphics[width=0.48\textwidth]{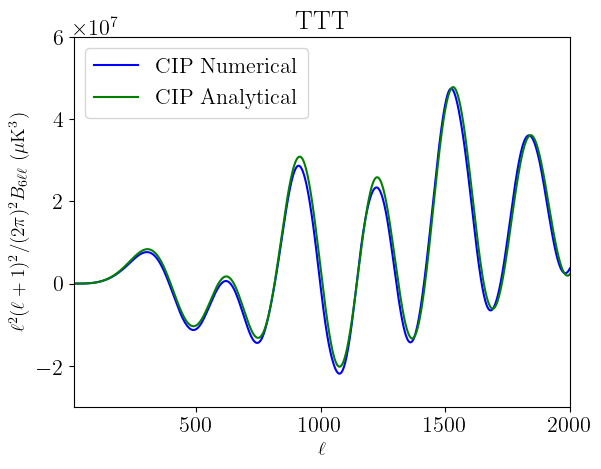}
		\includegraphics[width=0.50\textwidth]{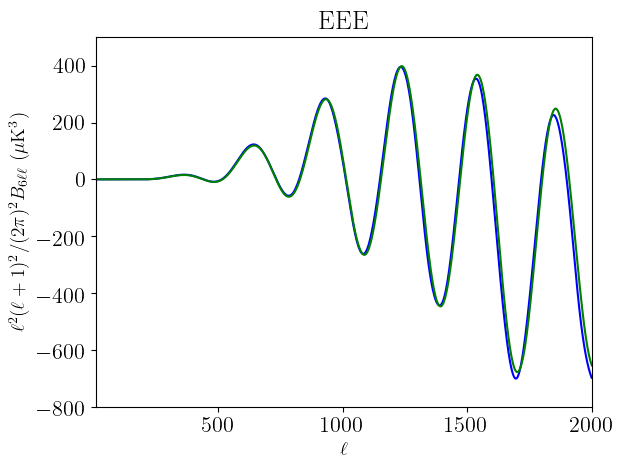}
    \caption{Bispectrum for a compensated isocurvature with $f_{\rm{CIP}}=-10^3$ (fully anti-correlated with the adiabatic mode) from the numerical calculation with SONG compared to the analytical formula, Eq.~\eqref{CIP_an_formula}, for TTT (left) and EEE (right).}
    \label{CIP_An}
\end{figure}

When correlated with the adiabatic mode, the CIP generates an enhancement to the bispectrum that is proportional to the root of the amplitude of the power spectrum of CIPs, as well as the correlation angle between the modes, i.e. it depends on the combination 
\be
f_{\text{CIP}}=\cos\theta_{\zeta\rm{CIP}}\sqrt{\frac{A_{\text{CIP}}}{A_s}}\,.
\ee
In the squeezed limit, an analytical result can be derived for this bispectrum using the results of Ref.~\cite{Smith:2017ndr} for the modulation of the temperature by CIPs, or the general result for squeezed bispectra of Refs.~\cite{Lewis:2011au,Lewis:2012tc}. The final result is given by
\be
\label{CIP_an_formula}
B_{\ell_L\ell_S\ell_S}^{X_1 X_2 X_3}=C_{\ell_L}^{X_1\Delta}\frac{\p C_{\ell_S}^{X_2 X_3}}{\p \bar\Delta}\,,
\ee
where $X_i=T,E$ depending on the case in question, $\Delta=\delta_b^0$ is the initial density contrast of baryons in the compensated mode, and $\bar\Delta$ is a perturbation to the background value of the baryon-to-dark matter ratio that keeps the total matter density fixed. It is also clear here that this bispectrum vanishes if there is no correlation between the CIP, $\Delta$, and the field, $X_1$. This happens unless $\Delta$ is correlated with $\zeta$, via $\Delta=f_{\mathrm{CIP}}\zeta^0+\Delta_{\rm{uncorr}}$. This implies that the first term above is
\be
C_{\ell_L}^{X_1\Delta}=f_{\text{CIP}} C_{\ell_L}^{X_1\zeta^0}\,.
\ee
Regarding the derivative term, this can be easily obtained by approximating the derivative via a finite difference, by simply calculating the spectrum $C_{\ell_S}^{X_2 X_3}$ for different values of $\bar\Delta$ and interpolating. This approximate result can then be used to test the accuracy of our modified version of SONG, as is shown in Fig.~\ref{CIP_An} for $TTT$ and $EEE$.

\begin{figure}[h]
    \centering
    		\includegraphics[width=0.49\textwidth]{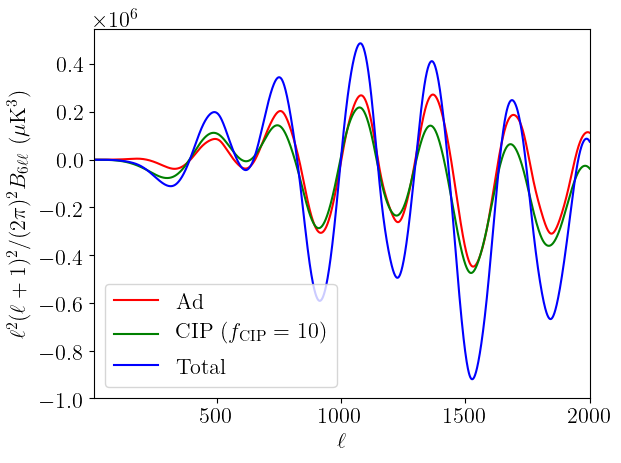}
		\includegraphics[width=0.49\textwidth]{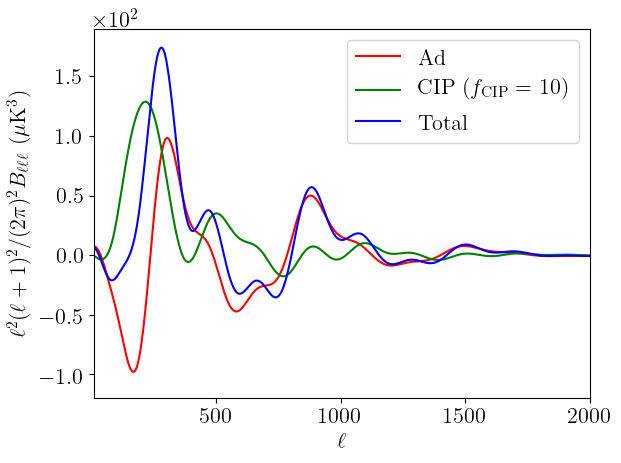}
		\includegraphics[width=0.50\textwidth]{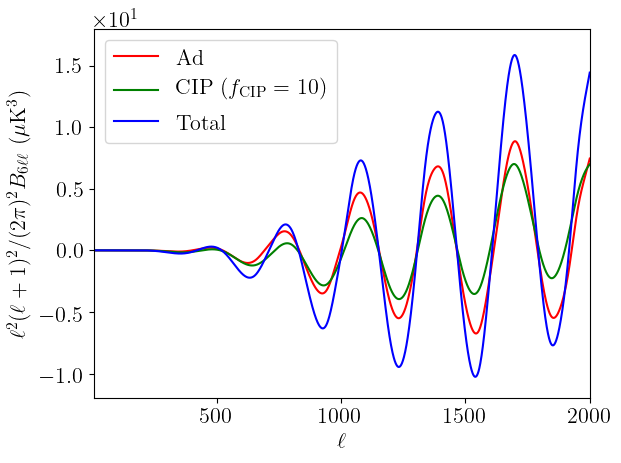}
		\includegraphics[width=0.49\textwidth]{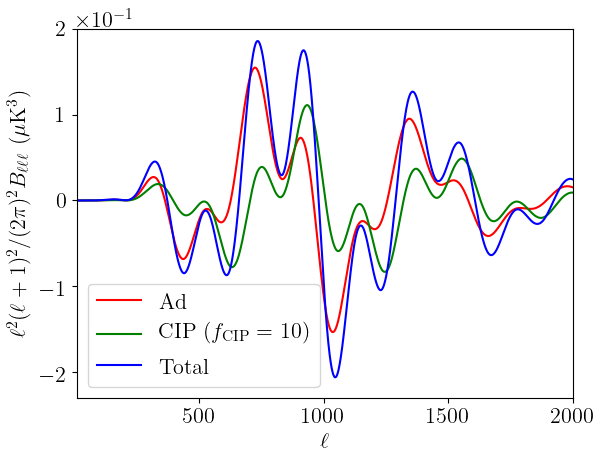}
    \caption{Bispectrum for a compensated isocurvature with $f_{\rm{CIP}}=10$ fully correlated with the adiabatic mode as compared to the adiabatic case for TTT (top row) and EEE (bottom row) in the squeezed (left) and equilateral (right) configurations.}
    \label{CIP_1}
\end{figure}

We can now present the main results of this section. In Fig.~\ref{CIP_1}, we show the resulting bispectrum for a fully correlated compensated isocurvature with $f_{\rm{CIP}}=10$. We can see that for this amplitude, the compensated isocurvature produces a contribution of similar size to the adiabatic mode, with similar shape in the squeezed limit, but clearly deviating from it in other configurations. This is again seen in the polarization case, but the size of the contribution from the compensated mode is slightly smaller.

The size of these signals is already relevant for probing models of inflation. A particular case is the curvaton scenario which is known to predict compensated modes. In certain versions of that model these isocurvatures can have amplitudes up to $f_{\rm{CIP}}=16.5$~\cite{He:2015msa}. Should the sensitivity of future CMB experiments become close to be limited only by cosmic variance and the forecasted signal-to-noise of $S/N=4$ be achieved for the intrinsic bispectrum~\cite{Pettinari:2014iha}, these contributions could be constrained, given that they are expected to approximately double the signal. In addition, for $f_{\rm{CIP}}=16.5$, our Fisher forecasts for the amplitude of the total intrinsic bispectrum reveal a signal-to-noise ratio of $S/N\approx 4$ already for $\ell_{\rm{max}}=2000$, when including both polarization and temperature in a cosmic variance limited experiment. While we use a Gaussian approximation for the covariance, this estimate should not change substantially beyond a factor of 2 in the worst case scenario. These results therefore suggest that a detection of this CIP mode from the curvaton model is possible via the analysis of the bispectrum in futuristic CMB experiments.

\begin{figure}[h]
    \centering
		\includegraphics[width=0.49\textwidth]{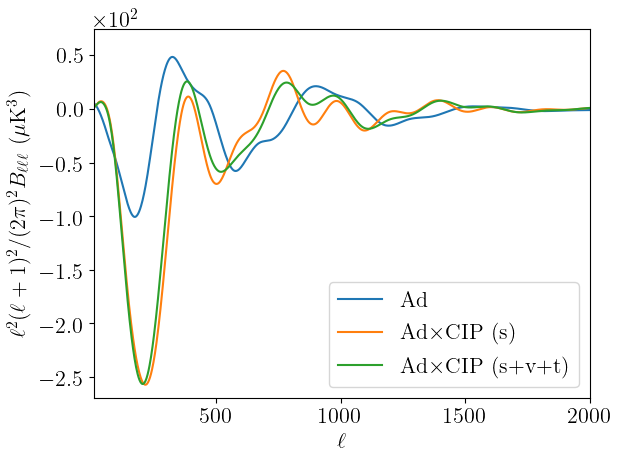}
    		\includegraphics[width=0.49\textwidth]{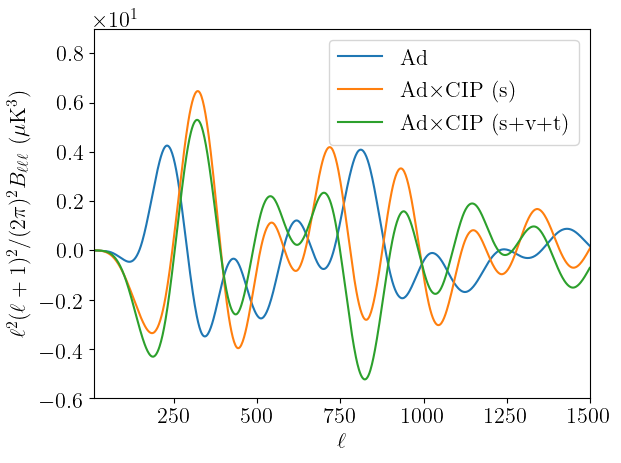}
    \caption{TTT (left) and ETT (right) equilateral bispectra for a compensated isocurvature with $f_{\rm{CIP}}=-20$ comparing the effects of scalar perturbations with the inclusion of vector and tensor modes. The adiabatic case with vector and tensor modes is included for context.}
    \label{CIP_svt}
\end{figure}

Before detailing our forecasts, we first study the influence of second-order vectors and tensors for the compensated isocurvature. In Fig.~\ref{CIP_svt} we show example bispectra in the equilateral configuration. We find that the effect of non-scalar modes is similarly small to the case of simple isocurvatures and again only seen in non-squeezed configurations, which motivates our choice of showing only the equilateral case. Given that the squeezed case dominates the signal by several orders of magnitude, the inclusion of non-scalar modes only negligibly increases the signal and does not affect our forecasts substantially, so we choose to neglect them in further calculations.

\begin{figure}[h]
    \centering
		\includegraphics[width=0.49\textwidth]{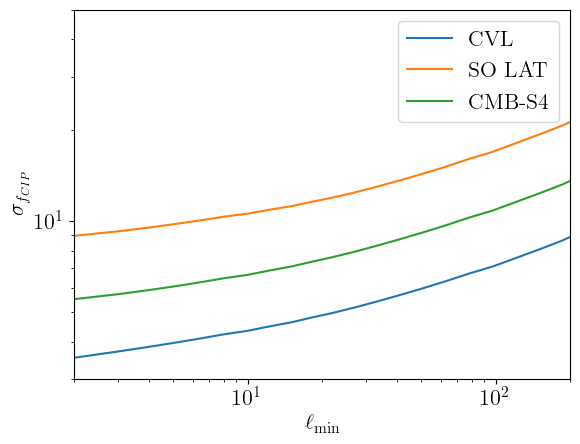}
    		\includegraphics[width=0.49\textwidth]{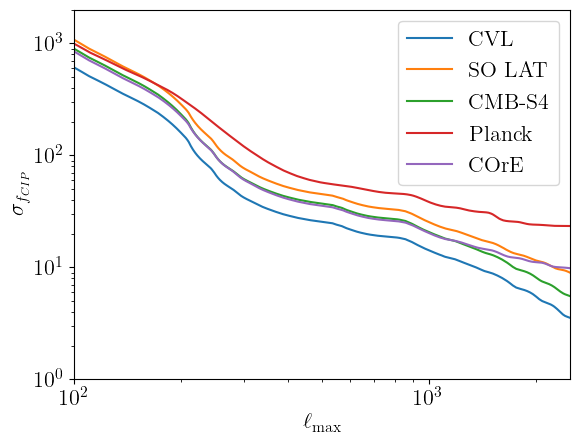}
    \caption{Forecasted uncertainty for the amplitude $f_{\rm{CIP}}$ as a function of $\ell_{\rm{min}}$, with fixed $\ell_{\rm{max}}=2500$ (left) and as a functions of $\ell_{\rm{max}}$ with fixed $\ell_{\rm{min}}=2$ (right).}
    \label{CIP_Fish}
\end{figure}

We now focus on the forecasted uncertainty for the amplitude $f_{\rm{CIP}}$ which we show in Fig.~\ref{CIP_Fish}.\footnote{The noise parameters for the experiments studied here are given in Refs.: \emph{Planck}~\cite{Planck:2013cta}, COrE~\cite{DiValentino:2016foa}, Simmons Observatory~\cite{Ade:2018sbj}, CMB stage 4~\cite{Abazajian:2019eic} and we use $f_{\rm{sky}}=0.65$ for the satellite experiments, $f_{\rm{sky}}=0.4$ for SO LAT and $f_{\rm{sky}}=0.5$ for CMB-S4.} We also summarise the results of the Fisher forecasts in table~\ref{Constraints}. We do not go above $\ell_{\rm{max}}=2500$ due to computational limitations, but most future experiments are expected to go beyond that, increasing the signal and achieving better constraints. At the same time, abandoning the Gaussian approximation for the covariance would increase the covariance and remove some constraining power. These simplifications are expected to cancel each other partially, but the values in table~\ref{Constraints} should be interpreted as approximate estimates of the real constraints possible from these experiments. In any case, it is clear that for larger $\ell_{\rm{max}}>2500$, ground-based experiments will provide better constraints due to the steeper scaling seen in the right plot of Fig.~\ref{CIP_Fish}. However, these ground-based experiments are not expected to reach $\ell\lesssim30$ and for that reason we also estimate the dependence on $\ell_{\rm{min}}$ for these experiments in the left plot of Fig.~\ref{CIP_Fish}, from which one can infer the degradation of constraints due to ignoring low $\ell$ data. For example, our estimate for the constraining power is expected to worsen by a factor 2 in the conservative scenario of ignoring $\ell\lesssim200$. Finally, we point out that we are ignoring atmospheric noise in our modelling, which is important at low $\ell$, so our estimates are still optimistic even in the cases in which ground-based experiments do measure fairly low $\ell$. In spite of all this, these issues would be effectively eliminated by using Planck data on large scales, which is already measured to extremely high accuracy and does not suffer from atmospheric noise. That is the scenario for which our results are the most accurate.

We can also see that the uncertainty on this CIP amplitude could be as low as $\sigma_{f_{\rm{CIP}}}\approx 3.5$ for an ideal experiment at $\ell_{\rm{max}}=2500$, whereas for \emph{Planck} our estimate is $\sigma_{f_{\rm{CIP}}}\approx 23$. Even with an error of $O(2)$ due to the Gaussian covariance approximation, these results are already comparable with the forecasts using the power spectrum of Ref.~\cite{He:2015msa}, which found $\sigma_{f_{\rm{CIP}}}\approx 22$ for \emph{Planck} and $\sigma_{f_{\rm{CIP}}}\approx 3$ for a cosmic variance limited (CVL) experiment. While it is expected that the signals in the power spectrum and bispectrum are correlated, a combination of both could improve constraints. This could be especially relevant given that the power spectrum is also sensitive to the uncorrelated CIP while the bispectrum is not, which would allow for a better separation of both effects. Other observables that also probe the uncorrelated mode, such as galaxy bias, could also be combined in this way to better separate both contributions.

\begin{table}
\centering
 \begin{tabular}{||c c c||} 
 \hline
 Experiment &\ \ &$\sigma_{f_{\rm{CIP}}}$  \\ [0.5ex] 
 \hline\hline
 Planck &\ \ & 23  \\ 
 \hline
 COrE &\ \ & 9.8  \\
 \hline
 SO &\ \ & 8.9  \\
 \hline
 CMB-S4 &\ \ & 5.5  \\
 \hline
 CVL &\ \ & 3.5  \\ 
 \hline
\end{tabular}
\caption{Fisher forecasts for $f_{\rm{CIP}}$ for different CMB experiments for $\ell_{\rm{max}}=2500$}
\label{Constraints}
\end{table}

\begin{figure}[h]
    \centering
    \includegraphics[width=0.49\textwidth]{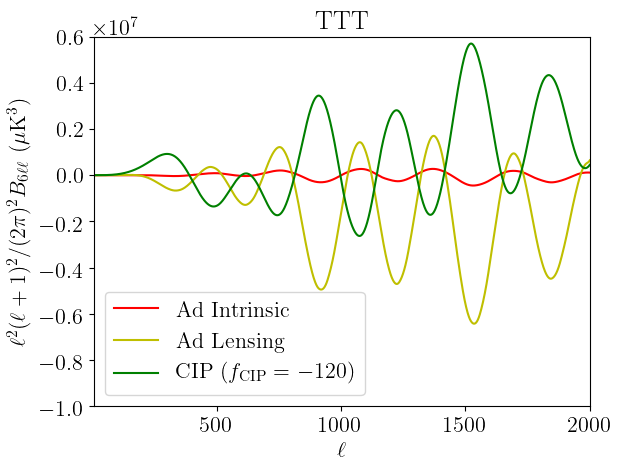}
    \includegraphics[width=0.48\textwidth]{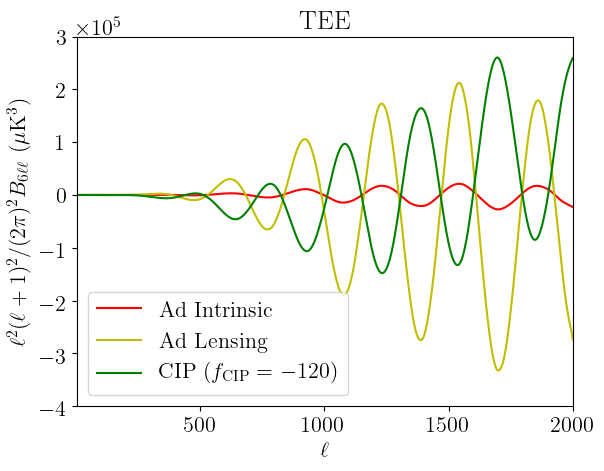}
    \caption{Bispectra for a compensated isocurvature with $f_{\rm{CIP}}=-120$ as compared to the adiabatic case including both the intrinsic contribution as well as that from lensing, for different combinations of $E$ polarization and temperature $T$ given by $X_1X_2X_3$ above each plot, with $X_1$ corresponding to the squeezed field.}
    \label{CIP_lensing}
\end{figure}

In spite of this signal not having been searched for in \emph{Planck} bispectrum data, it is already possible to use results based on that analysis to place constraints on CIPs. This is due to the known detection of the lensing contribution to the bispectrum, whose normalized amplitude has been measured to be $f^{\rm{lens}}_{\rm{NL}}=1.03\pm0.27$, taking into account both the temperature and polarization signals~\cite{Akrami:2019izv}. In Fig.~\ref{CIP_lensing}, we show two bispectra produced by CIPs compared to those generated by lensing in both TTT and TEE, where it is clear that their size is comparable. With this choice of sign, both contributions partially cancel, a situation which is clearly ruled out by the observation of the expected amplitude of the lensing bispectrum. In more detail, the bias on $f^{\rm{lens}}_{\rm{NL}}$ introduced by the compensated isocurvature for \emph{Planck} is approximately given by
\be
\Delta f^{\rm{lens}}_{\rm{NL}}\approx 2.7\times 10^{-3} f_{\rm{CIP}}\,,
\ee
whereas our forecasted error on the lensing bispectrum is $\sigma_{f^{\rm{lens}}_{\rm{NL}}}=0.27$, which is equal to that quoted by \emph{Planck}, given that the signal variance is neglected in both analysis. Under the same assumptions, this implies that we can place a constraint on $f_{\rm{CIP}}$, given by
\be
f_{\rm{CIP}}=1\pm100\,.
\ee
This constraint constitutes one of our main results and, while still not reaching forecasted constraints, it is already better than the current constraints on the amplitude of the uncorrelated CIP of $\sqrt{A_{\rm{CIP}}/A_s}\approx 500$. Future experiments could also use their analysis of the lensing bispectrum to further constrain this CIP amplitude. However, our forecasts indicate that a CVL experiment with $\ell_{\rm{max}}=2500$ would only reach $f_{\rm{CIP}}\approx 33$ using the bias on the lensing bispectrum, indicating that a full analysis of the CIP bispectrum would be more fruitful, potentially placing constraints an order of magnitude better. As mentioned above, both the Planck measurement and the signal-to-noise estimated here could be modified by the Gaussian approximation to the covariance, which could increase the errors by a factor of approximately 2.

Using the same approach for the local template would not improve constraints from \emph{Planck} as this bias would only be detectable for $f_{\rm{CIP}}\approx 150$. However, this changes for a CVL experiment, which would see $f^{\rm{local}}_{\rm{NL}}$ biased by $1\sigma$ for $f_{\rm{CIP}}\approx 10$, which is not only promising for probing CIPs but also indicates that ignoring these isocurvature modes could lead to a false detection of local non-Gaussianity in future experiments. An example of this is the curvaton scenario with $f_{\rm{CIP}}=16.5$, for which our forecasts indicate that the bias on local non-Gaussianity could reach $\Delta f^{\rm{local}}_{\rm{NL}}=2.5$ for a CVL experiment. This is the same order of magnitude as the non-Gaussianity expected from that curvaton scenario ($f^{\rm{local}}_{\rm NL}\approx 6$ \cite{He:2015msa}) and would thus crucially change the interpretation of the data, should non-Gaussianity of this order of magnitude be detected in the future.\\

\begin{figure}[h]
    \centering
		\includegraphics[width=0.49\textwidth]{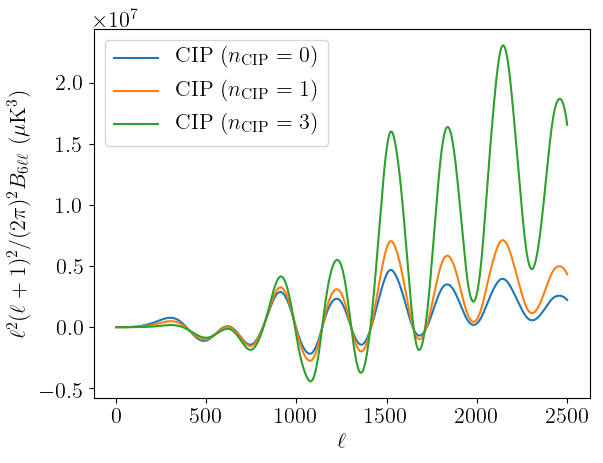}
    		\includegraphics[width=0.5\textwidth]{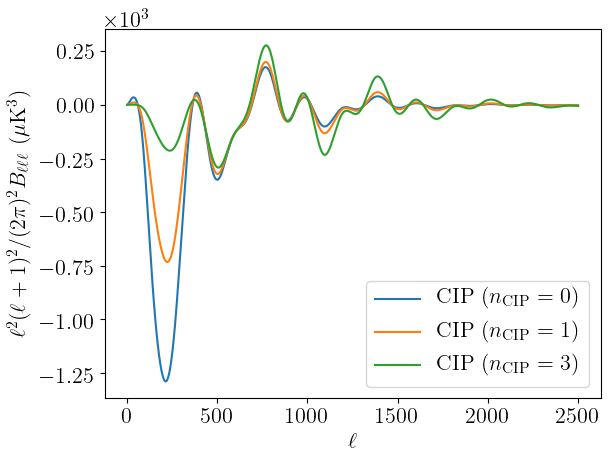}
    \caption{TTT bispectra for compensated isocurvatures with varying spectral index in the squeezed (left) and equilateral (right) configurations.}
    \label{CIP_varn}
\end{figure}

Finally, we explore scenarios in which the compensated isocurvature has a different spectral index than the adiabatic mode. While this is not known to be possible in curvaton-like scenarios, it can be achieved phenomenologically if the relation between the CIP and $\zeta$ is scale dependent, i.e. $\Delta=f_{\mathrm{CIP}}(k)\zeta^0$ with 
\be
f_{\mathrm{CIP}}(k)=\bar{f}_{\mathrm{CIP}}\left(\frac{k}{k_*}\right)^{(n_{\rm{CIP}}-n_s+1)/2}\,.
\ee
We show the resulting bispectra for various $n_{\rm{CIP}}$ in Fig.~\ref{CIP_varn}. As the spectral index increases, it is clear that the bispectrum is reduced on large scales in both configurations shown, but drastically increases on small scales, specially in the squeezed configuration, an effect which is also seen in polarization.

This amplification of the signal on small scales makes these scenarios easier to constrain and indeed we find the signal-to-noise of this bispectrum increases substantially for large spectral indices. We forecast that constraints on $\bar{f}_{\mathrm{CIP}}$ can be up to 5 times tighter for $n_{\rm{CIP}}=3$ and are improved by a factor $5/3$ for $n_{\rm{CIP}}=1$.  The bias on the lensing and local templates are also increased by factors of 4 and 6.5, respectively for $n_{\rm{CIP}}=3$, which could once more be used to probe these isocurvature modes. All of this implies that the bispectrum of the CMB can be used to constrain both the amplitude and the spectral index of correlated compensated isocurvature modes.

\section{Conclusions}
\label{sec:conc}

We have studied the effects of isocurvature modes on the intrinsic bispectrum of the CMB. This represents the first calculation of this bispectrum for all types of isocurvatures, thus shedding light on how it is affected by general initial conditions.

Our results show that the bispectra sourced by pure isocurvature modes are substantially smaller than that created by the adiabatic mode. When mixed, the combination of adiabatic and single isocurvature modes gives rise to substantial variations in the total bispectrum, which resemble the effects of local primordial non-Gaussianity. While interesting in their own right, these effects are found to be too small for realistic amplitudes of the isocurvature modes and therefore would not be useful in constraining them beyond what has already been achieved by \emph{Planck}. We also test more complex isocurvature modes, including a correlation with the adiabatic mode as well as variations of spectral index. We find that these changes affect the signal substantially, but not enough to make these complex scenarios more likely to be probed by their effects on the intrinsic bispectrum.\\

Furthermore, we study the compensated isocurvature mode, whose effects are substantially larger, given its potentially large amplitude. After showing the validity of our modified version of the Boltzmann solver SONG, we demonstrate the possibility of observations of the bispectrum to constrain the amplitude of the compensated mode when it is correlated with the adiabatic mode. We show that for a relative amplitude of $f_{\mathrm{CIP}}=10$, the bispectrum sourced by the CIP is similar in amplitude to the intrinsic bispectrum generated by the adiabatic mode, while for an amplitude of $f_{\mathrm{CIP}}=100$ it is similar in size to the lensing contribution detected by \emph{Planck}. 

We use the Fisher matrix formalism with the Gaussian covariance approximation to forecast the observability of the CIP contribution to the bispectrum for different CMB experiments, as well as to what extent its neglect biases measurements of other bispectrum signals. We estimate that a dedicated analysis of \emph{Planck} bispectrum data, using e.g. the modal~\cite{Fergusson_2010} or binned~\cite{Bucher:2015ura} estimators, could already place competitive constraints on $f_{\mathrm{CIP}}$, comparable to those attainable from an analysis of the power spectrum. We further show that future experiments, such as COrE, SO and CMB-S4, could improve on those constraints below $f_{\mathrm{CIP}}=10$ and thus allow for probing the particular curvaton scenario that predicts $f_{\mathrm{CIP}}=16.5$. We summarise these results in table~\ref{Constraints}. 

In addition, we find that even without a dedicated search for the CIP bispectrum, interesting constraints can be achieved by analysing other bispectrum templates, such as the lensing and local shapes, and estimating the bias caused on those searches when ignoring the CIP contribution. From this we find that \emph{Planck} already places a constraint of $f_{\rm{CIP}}=1\pm100$ given the non-observation of an anomalous lensing bispectrum. We also find that future experiments could use the lack of detection of local non-Gaussianity to probe $f_{\rm{CIP}}$ of $O(10)$. Furthermore we conclude from this that substantial biases can occur in the measurement of primordial non-Gaussianity from ignoring this CIP contribution, which could be crucial for the correct interpretation of a future detection.

We also vary the spectral index of the compensated isocurvature, finding that a bluer spectrum would result in stronger constraints on the amplitude. We therefore conclude that both spectral parameters can be probed via the bispectrum.

In conclusion, isocurvature modes leave distinct imprints on the intrinsic bispectrum of the CMB and its analysis can be used to constrain the compensated isocurvature mode.

\acknowledgments

PC was supported by STFC grant ST/P000592/1 and by a UKRI Future Leaders Fellowship, grant MR/S016066/1, and KAM is supported by STFC grant ST/P000592/1.
The authors are grateful to David Mulryne, Cyril Pitrou, William Coulton and Atsuhisa Ota for useful discussions.

\bibliographystyle{JHEPmodplain}
\bibliography{biblio_mag}

\end{document}